\documentclass[10pt,a4paper]{article}

\usepackage[T1]{fontenc}
\usepackage[utf8]{inputenc}
\usepackage{lmodern}
\usepackage{microtype}
\usepackage[margin=2.4cm]{geometry}
\usepackage{amsmath,amssymb,amsthm}
\usepackage{graphicx}
\usepackage{booktabs}
\usepackage{setspace}
\usepackage[labelfont=bf,font=footnotesize]{caption}
\usepackage[numbers,sort&compress]{natbib}
\usepackage{xcolor}
\usepackage{xurl}
\usepackage[colorlinks=true,linkcolor=black,citecolor=blue!50!black,
            urlcolor=blue!50!black]{hyperref}

\usepackage{mathtools}
\usepackage{bm}
\usepackage{units}
\usepackage{enumitem}
\graphicspath{{figures/}}

\newcommand{\PaperTitle}{A reduced formulation for the free-wave propagation analysis in composite structures}
\newcommand{\PaperAuthors}{C. Droz$^{1,2}$, J.-P. Lain\'e$^{1}$, M.N. Ichchou$^{1}$ and G. Inqui\'et\'e$^{2}$}
\newcommand{\PaperAffil}{%
  $^{1}$Ecole Centrale de Lyon, 36 Avenue Guy de Collongue, 69134 Ecully Cedex, France\\
  $^{2}$Airbus Helicopters, A\'eroport Inter. Marseille-Provence, 13725 Marignane Cedex, France}
\newcommand{\PaperDate}{Accepted manuscript, originally 2014}
\newcommand{\PaperKeywords}{Wave propagation, Composite waveguides, Dispersion curves, Reduction, Wave finite elements}

\newcommand{\PaperDOI}{10.1016/j.compstruct.2014.03.017}
\newcommand{\HalID}{hal-03372953}
\newcommand{\ArxivID}{}          

\newcommand{\CiteAs}{%
  C. Droz, J.-P. Lainé, M.N. Ichchou, G. Inquiété, \emph{\PaperTitle}, Composite Structures 113 (2014) 134-144, doi:10.1016/j.compstruct.2014.03.017}

\makeatletter
\newcommand{\HasDOI}{%
  \ifx\PaperDOI\@empty\expandafter\@secondoftwo\else\expandafter\@firstoftwo\fi}
\newcommand{\HasArxiv}{%
  \ifx\ArxivID\@empty\expandafter\@secondoftwo\else\expandafter\@firstoftwo\fi}
\makeatother

\providecommand{\doi}[1]{\href{https://doi.org/#1}{doi:#1}} 

\begin{document}

\begingroup
\renewcommand{\thefootnote}{\fnsymbol{footnote}}
\begin{center}
{\LARGE\bfseries \PaperTitle\par}
\vspace{0.7em}
{\normalsize \PaperAuthors\footnotemark[1]\footnotemark[2]\par}
\end{center}
\footnotetext[1]{\mbox{}\\[-0.7em]\raggedright\PaperAffil}
\footnotetext[2]{\PaperDate.
  Author's version with AI-assisted typesetting and layout. Not the publisher version of record.
  \HasDOI{Definitive publication: \doi{\PaperDOI}.}{No journal DOI yet.}
  HAL: \href{https://hal.science/\HalID}{\HalID}.
  \HasArxiv{arXiv: \href{https://arxiv.org/abs/\ArxivID}{\ArxivID}.}{}}
\endgroup
\setcounter{footnote}{0}

\vspace{0.85em}
{\small
\noindent\textit{How to cite this version.}\\[0.35em]
\raggedright \CiteAs
\par}
\vspace{1.05em}

\begin{abstract}
\noindent
This paper presents an improved numerical strategy for the broadband analysis of wave propagation in composite or complex cross-sectional waveguides using the wave finite element method (WFE). Numerical analysis of such structures requires highly discretized finite element models and leads to extensive computations. The proposed formulation relies on a projection of the cross-sectional transfer matrices on a reduced set of shape functions associated to propagating waves. Dispersion curves are then predicted only using a reduced number of eigenvectors. The performances and stability of this method are evaluated using the wavenumbers and wave shapes. Validations are provided for a sandwich composite beam and a cylindrical elasto-acoustic waveguide.
\end{abstract}

\noindent\textit{Keywords:} \PaperKeywords


\section{Introduction}
Wave propagation in composite waveguides in a broadband frequency range is widely investigated in automotive and aerospace industry. A waveguide is a structure whose main dimensions exhibit a periodicity or homogeneity in such a way that the propagation of mechanical energy in the main direction (axis of a beam or plane of a plate) is privileged. Waveguide hypothesis can significantly reduce the size of the problem, since the behavior of a single sub-structure of the guide yields the response of the entire structure. Dynamical behavior of such structures is determined by evaluating the set of structural waves propagating through a cross-section \cite{Duhamel2006,Mead2009}. One of the major interests of guided waves is their potential for travelling long distances at velocities governed by the dispersion phenomena. The knowledge of these dispersion properties for propagating waves is fundamental for an effective use in engineering, for example in the field of structural health monitoring (SHM).\\
Numerical prediction of these different waves and their dispersion curves has been extensively studied in last decades. The semi-analytical finite element (SAFE) and wave and finite element (WFE) methods are, among others, very efficient tools for this purpose. In the SAFE method, sinusoidal functions are employed to formulate the displacement field in the direction of propagation. Nevertheless, it is necessary to develop specific semi-analytical elements for each application, which can severely limit its interest for industrial purposes. In order to overcome these limitations, the WFE method combines periodic structure theory (PST) introduced in Mead \cite{Mead1996} with a finite element method (FEM). \\
Therefore conventional finite element software packages can be easily used to compute mass and stiffness matrices of the whole structure. The one-dimensional WFE method was successfully applied to a wide range of waveguides as beam-like structures \cite{Thompson1993,Mencik2005,Mace2005}, plates \cite{Ichchou2008} and more complex geometries as thin-walled structures \cite{Houillon2005}, tyres \cite{Waki2009a}, pipes \cite{Bareille2012} and curved layered shells \cite{Chronopoulos2014}.\\
As the application field of WFE method reaches structurally advanced composite structures, various numerical difficulties can appear, especially for one-dimensional formulation which involves larger cross-sections. Poor-conditioning of the transfer matrix can lead to numerical errors (see Zhong and Williams \cite{Zhong1995} for alternative formulations), aliasing effects and round-off errors can also appear if cross-section length is not sized carefully. However, for the determination of propagating waves in industrial waveguides involving a large number of degrees of freedom, major obstacle remains the large CPU time needed to solve the eigenproblem. Some numerical issues were investigated for example by Waki et al. \cite{Waki2009b} and a reduction strategy based on a contributing waves selection was proposed by Mencik \cite{Mencik2012} to compute forced response of elastic waveguides \cite{Reno2010}. Mencik et al. proposed a substructuring technique to compute the appropriate wave motion in multi-layered waveguides. Homogenisation techniques were also investigated \cite{Gopalakrishnan2008} to apply the WFE to laminated composites.\\

Although the aforementioned reduction techniques are interesting to compute forced responses using a reduced number of wave modes, they do not reduce the numerical costs associated with the computation of the wave basis. Therefore, these techniques do not provide a general reduction strategy to compute the propagating waves in elaborated waveguides. This paper presents a method, based on classical formulation of the periodic structure theory, to calculate the dispersion curves of propagating waves for complex cross-section or composite waveguides involving an important number of degrees of freedom. This alternative formulation of WFE method relies on a frequency interpolation of the transfer matrix eigenvectors through a subset of eigensolutions. 
Thus, the propagating waves can be determined accurately solving a smaller eigenproblem, enabling the application of the WFE method to a wide range of sophisticated cross-sectional waveguide configurations.\\

The paper is organized as follows. In Section \ref{wfe_classique}, a brief overview of the classical WFE formulation is shown. Section \ref{wfe_reduite} describes the proposed reduction strategy. The model order reduction is formulated for the spectral problem and the strategy of wave interpolation is described; the wave basis is defined next, using a reduced set of propagating waves computed at the cut-on frequencies, associated to the appearance of new propagating waves over the frequency band; a method is then proposed to improve the basis orthogonality and approximate eigenvectors between the cut-on frequencies. Numerical examples are brought in Section \ref{numerical}. The first application concerns a three-layered sandwich beam; both the classical WFE formulation and an analytical low frequency solution described in \cite{Banerjee2005,Mencik2008} are discussed, and the requirement for a refined FEM of cross-section is highlighted; the reduction strategy is then applied to a detailed FEM model. In the second example, the reduced WFE formulation is extended to an elasto-acoustic problem; dispersion curves are computed for a cavity filled with fluid and compared to the analytical solution. 

\section{Overview of the WFE\label{wfe_classique}}
\subsection{Free wave propagation in 1D-waveguides}
\begin{figure}[!ht]
  \centering
   \includegraphics[width=0.7\columnwidth]{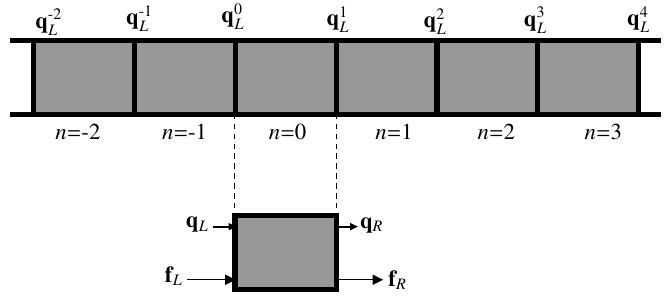}
   \caption{Illustration of a waveguide and the state vector of a unit cell \cite{Duhamel2006}.}
   \label{cell_wfe}
\end{figure}

In this section, a formulation of the WFE method is given for free wave propagation in a one-dimensional straight elastic and dissipative waveguide. The structure can be assimilated to N identical subsystems of length d connected along the main direction $x$. A unit cell of the waveguide is illustrated Fig. \ref{cell_wfe}.
Displacements and forces are written as \textbf{q} and \textbf{f}, and subscripts '$L$' and '$R$' denote the left and right edges of a cell. Both edges have the same number n of degrees of freedom. Mesh compatibility is assumed between the N subsystems. The discrete dynamic equation of a cell at frequency $\omega$ is given by:
\begin{equation}
 (-\omega^2\mathbf{M}+j\omega \mathbf{C}+\mathbf{K})\mathbf{q}=\mathbf{f}
\end{equation}
where $\mathbf{M}$,$\mathbf{C}$,$\mathbf{K}$ are the mass, damping and stiffness matrices, respectively. For periodic structures, condensation on the left and right cross-sections of the inner DOF's is required. Introducing the condensed dynamic stiffness operator $\mathbf{D}=-\omega^2\mathbf{M}+j\omega \mathbf{C}+\mathbf{K}$ and reordering degrees of freedom, equation can be stated as follows:
\begin{equation}
 \left[ \begin{array}{cc}
  \mathbf{D}_{LL} & \mathbf{D}_{LR} \\
  \mathbf{D}_{RL} & \mathbf{D}_{RR}
 \end{array}\right] 
\left\lbrace \begin{array}{c}
  \mathbf{q}_{L} \\
  \mathbf{q}_{R}
 \end{array}\right\rbrace 
 = \left\lbrace \begin{array}{c}
  \mathbf{f}_{L} \\
  \mathbf{f}_{R}
 \end{array}\right\rbrace 
 \label{eqmotion}
\end{equation}
where $\mathbf{D}_{LL}$ and $\mathbf{D}_{RR}$ are symmetric and $\mathbf{D}^t_{LR}=\mathbf{D}_{RL}$. Denoting $\lambda=e^{-j\kappa d}$ the propagation constant describing wave propagation over the cell length $d$ and $\kappa$ associated wavenumber, considering force equilibrium $\lambda \mathbf{f}_L+\mathbf{f}_R=0$ in a cell and invoking Bloch's theorem \cite{Brillouin1953}, $\mathbf{q}_R=\lambda \mathbf{q}_L$ into Eq.\,(\ref{eqmotion}) leads to the following quadratic spectral problem \cite{Duhamel2006}:
\begin{equation}
( \lambda_i \mathbf{D}_{LR} + \frac{1}{\lambda_i} \mathbf{D}_{RL} + \mathbf{D}_{LL} + \mathbf{D}_{RR}) \mathbf{\Phi}^\mathbf{q}_i = 0
\label{eqspectral}
\end{equation}
where $((\mathbf{\Phi}^q)_i,\lambda_i)_{i=1,...,2n}$ stands for the wave modes of the waveguide. The associated eigenvalue problem can be formulated by an appropriate state vector $\mathbf{\Phi}^q=[(\mathbf{\Phi}^q)^t , (\mathbf{\Phi}^f)^t]^t$, leading to a symplectic transfer matrix $\mathbf{T}$.
\begin{equation}
 \mathbf{T} \left\lbrace \begin{array}{c}
  \mathbf{\Phi}^q \\
  \mathbf{\Phi}^f
 \end{array}\right\rbrace 
 = \lambda\left\lbrace \begin{array}{c}
  \mathbf{\Phi}^q \\
  \mathbf{\Phi}^f
 \end{array}\right\rbrace 
 \label{eqeigen}
 \end{equation}
 with
 \begin{equation}
 \mathbf{T} = \left[\begin{array}{cc}
  \mathbf{D}_{LR}^{-1} \mathbf{D}_{LL} & \mathbf{D}_{LR}^{-1} \\
  \mathbf{D}_{RL} - \mathbf{D}_{RR} \mathbf{D}_{LR}^{-1} \mathbf{D}_{LL} & - \mathbf{D}_{RR} \mathbf{D}_{LR}^{-1}
 \end{array}\right]
\end{equation}
where eigenvectors represent both nodal displacements and forces associated to a wave mode. The dynamical behavior of the global system can be expressed by expanding amplitudes of incident and reflected waves on a basis of eigenvectors. If the structure is undamped, solutions are divided into propagative waves, whose wavenumbers are real, and evanescent waves for which wavenumbers are imaginary. In dissipative case, complex wavenumbers are associated to decaying waves.

\subsection{Computational issues for a complex cross-section}
In practice, direct computation of the eigenproblem Eq.\,(\ref{eqeigen}) can be prone to numerical errors when the meshed cross-section involves a large number of degrees of freedom. Indeed, the transfer matrix T requires to inverse matrix $\mathbf{D}_{LR}^{-1}$ which can be poorly conditioned. To limit this issue, various formulations of the eigenproblem are available, for example:
\begin{equation}
 \left[\begin{array}{cc}
  -\mathbf{D}_{RL} & -(\mathbf{D}_{LL}+\mathbf{D}_{RR}) \\
  \mathbf{0} & -\mathbf{D}_{RL}
 \end{array}\right]
 \left\lbrace \begin{array}{c}
  \mathbf{q}_L \\
  \lambda \mathbf{q}_L
 \end{array}\right\rbrace 
 = \lambda \left[\begin{array}{cc}
  \mathbf{0} & \mathbf{D}_{LR} \\
  -\mathbf{D}_{RL} & \mathbf{0}
 \end{array}\right] 
 \left\lbrace \begin{array}{c}
  \mathbf{q}_L \\
  \lambda \mathbf{q}_L
 \end{array}\right\rbrace 
 \label{alternative_eigeq}
\end{equation}
However, numerical errors can become serious when eigenvalues are either very large ($\lambda_i$) or very small ($1/\lambda_i$), see Zhong and Williams \cite{Zhong1995} for a detailed discussion. When complex waveguides are considered, an insufficient discretization of the cross-section will produce significant errors, especially for eigensolutions associated to waves whose section shape have a short wavelength, whereas refined meshes exhibit numerous evanescent solutions, thus considerably increases computation time and worsen round-off errors due to the truncation of inertia terms, see Waki \cite{Waki2009b}.\\

Yet, structures considered in this paper require a high degree of precision due to their geometry, their inner components or for high order wave shapes calculation. In these situations computation time grows exponentially with the number of nodes involved. Classical techniques based on modal basis reduction are not available, since the cross-section boundary conditions are arbitrary for a uniform waveguide or subjected to structural periodicity otherwise. Such an issue is addressed in the next section, introducing projection on a reduced set of shape functions.

\section{Proposed reduction strategy\label{wfe_reduite}}
\subsection{Spectral problem projection\label{3.1}}
\subsubsection{Definition of a subspace for model order reduction}
The basic principle of Model Order Reduction (MOR) is to approximate Eq.\,(\ref{eqeigen}) by creating a low order system over a specified range of frequencies. Spectral problem (\ref{eqspectral}) can be rewritten:
\begin{equation}
 \mathbf{S}(\lambda,\omega)\mathbf{u}_L(\omega)=0
 \label{spectral_system}
\end{equation}
where S stands for:
\begin{equation}
 \mathbf{S}(\lambda,\omega)=\lambda \mathbf{D}_{LR}(\omega)+(\mathbf{D}_{LL}(\omega)+\mathbf{D}_{RR}(\omega))+\frac{1}{\lambda}\mathbf{D}_{RL}(\omega)
\end{equation}
Most MOR techniques can be interpreted as the formulation of two rectangular matrices $\mathbf{B}_{(r,n)}$ and $\mathbf{A}_{(n,r)}$ with $r \ll n$, which are used for system (\ref{spectral_system}) linear projection:
\begin{equation}
 \mathbf{\tilde{S}}_{(r,r)}=\mathbf{B}^T \mathbf{S}_{(n,n)} \mathbf{A}
 \label{matrices_A_B}
\end{equation}
The quality of this approximation strongly depends on the subspaces spanned by these matrices. Although matrices $\mathbf{A}$ and $\mathbf{B}$ can be defined numerous ways, numerical stability commonly requires $\mathbf{A}$ to be real, equal to $\mathbf{B}$ and orthogonal \cite{Schilders2008}:
\begin{equation}
 \mathbf{A}=\mathbf{B}, \quad \mathbf{B}\in\mathbb{R}^{n*r} \quad\textrm{and}\quad  \mathbf{B}^T \mathbf{B}=\mathbf{I}
 \label{stability}
\end{equation}
An effective way to obtain a reliable and low order reduced system is to make use of solution subsets from the initial system. This process is used, among others, for modal expansion or Krylov subspaces reduction \cite{Watkins2007}.

\subsubsection{Modal shapes - wave shapes\label{sec_wavemode}}
All reduced subspaces are not equivalent. In the field of structural analysis, modal analysis provides an efficient tool for numerous model order reduction strategies. Ritz method is based on the hypothesis that displacements field of a structure is spanned by its modal shapes $\left\lbrace \mathbf{\phi}_i\right\rbrace _{i=1,...,\infty}$. In a wave approach context, displacement of a bounded structure under harmonic excitation can be expressed as a sum of wave mode amplitudes \cite{Mencik2005}. Each stationary mode being linked to a wave reflection on the boundaries, both approaches can theoretically lead to the same conclusions \cite{Langley1997}. Stationary modes are boundary dependent and modal approaches can lead to important system reduction if the structure has a low modal density, but suffer limitations in mid-frequencies. Moreover, modal selection has numerous disadvantages, among other, a limited selection criteria: each mode is associated to a wave at a specific boundary dependent frequency.
\begin{figure}[!ht]
  \centering
   \includegraphics[width=0.6\columnwidth]{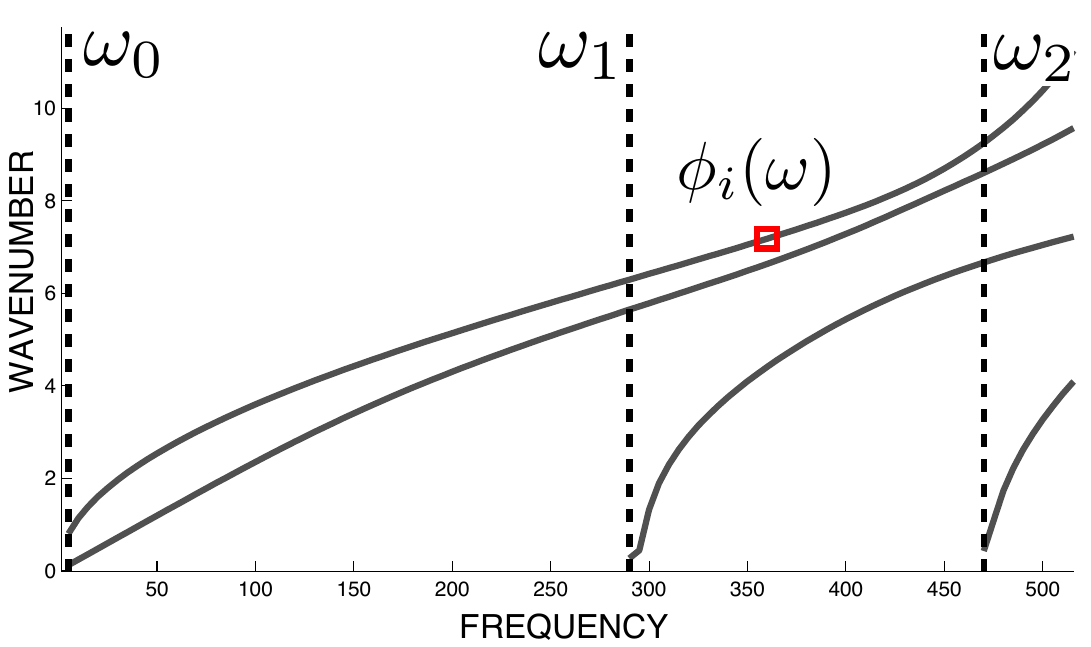}
   \caption{Illustration of the dispersion curves used for the eigenvectors approximation.}
   \label{illustr_red}
\end{figure}
Consider the dispersion curves shown Fig. \ref{illustr_red}. A limited set of eigenvectors is retained at frequencies $\omega_0$, $\omega_1$ and $\omega_2$, supposing an appropriate selection criteria can be found. They are denoted $\left\lbrace \vec{\phi}_i^{(0)}\right\rbrace _{1\leq i\leq r_0}$, $\left\lbrace \vec{\phi}_i^{(1)}\right\rbrace _{1\leq i\leq r_1}$ and $\left\lbrace \vec{\phi}_i^{(2)}\right\rbrace _{1\leq i\leq r_2}$. Considering that these eigenvectors represent the waves shapes through an elementary section, the wave shape $\vec{\phi}(\omega)$ shown Fig. \ref{illustr_red} can be approximated as:
\begin{equation}
 \vec{\phi}(\omega) \simeq \sum_{i=1}^{r_0} \vec{\phi}_i^{(0)} \nu_i^{(0)}(\omega) + \sum_{i=1}^{r_1} \vec{\phi}_i^{(1)} \nu_i^{(1)}(\omega) +  \sum_{i=1}^{r_2} \vec{\phi}_i^{(2)} \nu_i^{(2)}(\omega)
\end{equation}
where $\nu_i^{(0)}$, $\nu_i^{(1)}$, $\nu_i^{(2)}$ are the reduced wave amplitudes. Eigenvectors $\vec{\phi}_i$ are associated to propagation constants $(\lambda_i,1/\lambda_i)$ with the convention $|\lambda_i|\leq1$ for positive-going solutions and $|\lambda_i|\geq1$ for negative-going solutions. Their norms are closer to unity for propagating waves and smaller for evanescent waves, in the dissipative case, $|\lambda|<1$. These eigenvalues provide a simple criteria to sort and filter propagating waves (details are given in Section \ref{sec_lambda})). The following section describes the selection process and a method for computing an appropriate set of eigenvectors.

\subsection{Definition of the wave basis}  
\subsubsection{Selecting the appropriate waves}
As noticed Section (\ref{sec_wavemode}), accurate approximations require the exact solution to be a linear combination of the chosen eigenvectors. When a wave shape remains the same on a given frequency band, eigensolutions associated to this wave can be interpolated using an single, chosen eigenvector. This interpolation strategy is based on the hypothesis that major variations in the deformed shapes of propagating waves occur close to their cut-on frequencies.\\

This is an important statement for eigenvectors interpolation. Therefore, computation of the exact solutions can be limited to a reduced set of chosen frequencies, where eigenvectors are prone to important variations. Singular deformed shape can be observed in several situations. First, when the wave enters the propagative domain the new mode of propagation is generally associated with a modification of its wave shape. In higher frequency, boundary effects can occur. Other phenomena as coupling effects, which are due to energetic exchanges between wave modes, can also contribute to waves variations. Waves shapes at cut-on frequencies are expected to be good candidates to enrich the reduced basis, supposing a effective determination of their frequencies can be done. 

\subsubsection{Identification of progressive waves appearance\label{sec_lambda}}
Propagating waves have a positive wavenumber and low imaginary part, excepted for highly decaying waves. On the other hand, they are evanescent and have a negative wavenumber before they reach their cut-off frequencies. Considering wavenumbers are, among other, a continuous function of the frequency, the existence of a value $\omega_A$ for which the wavenumber $\kappa(\omega_A)=0$ is assured. In terms of propagation constants, this means $\lambda=1$. The frequencies $\omega_A$ associated to eigenvalues $\lambda=1$ can be determined by rewriting the quadratic eigenvalue problem Eq.\,(\ref{spectral_system}):
\begin{equation}
 \mathbf{S}(1,\omega_A)\mathbf{\Phi}_A=0
\end{equation}
Wave cut-on frequencies and associated wave shapes $\mathbf{\Phi}_A$ are then solutions of the eigenproblem of size $n$:
\begin{equation}
 (\mathbf{K}^\ast_{LL}+\mathbf{K}^\ast_{RR}+\mathbf{K}^\ast_{RL}+\mathbf{K}^\ast_{LR})\mathbf{\Phi}_A = \omega_A^2(\mathbf{M}_{LL}+\mathbf{M}_{RR}+\mathbf{M}_{RL}+\mathbf{M}_{LR})\mathbf{\Phi}_A
 \label{eqappearance}
\end{equation}
where $\mathbf{K}^\ast=(1+j\eta)\mathbf{K}$ is the complex stiffness considering Rayleigh damping ($\alpha=0$, $\beta=\eta/\omega$).\\

Let us denote $n_0$ the number of first-order waves, $n_2$ the number of higher-order waves in $\Delta\omega$ and $n_A$ the number of cut-on frequencies determined using Eq.\,(\ref{eqappearance}). As the first-order waves have the same cut-on frequency (at $\omega = 0$), the total number of cut-on frequencies is $n_A = 1 + n_2$. This means that the number of cut-on frequencies is lower than the total number: $n_0 + n_2$ of propagating waves in the frequency range.

\subsubsection{Exact wave computation using classical WFEM}
Denoting $\omega_{A(k)}$ with $1\leq k\leq n_A$ the cut-on frequency of the $k$th wave, the exact eigensolutions can be computed for each new propagating wave from the system of $n_A$ independent quadratic eigenproblems:
\begin{equation}
\left\lbrace\begin{array}{c}
 \mathbf{S}(\lambda,\omega_{A(1)}) \left\lbrace \mathbf{\Phi}^{u_L}\right\rbrace 
= \left\lbrace  \mathbf{0}  \right\rbrace \\
...\\
 \mathbf{S}(\lambda,\omega_{A(k)}) \left\lbrace \mathbf{\Phi}^{u_L}\right\rbrace 
= \left\lbrace  \mathbf{0}  \right\rbrace \\
...\\
 \mathbf{S}(\lambda,\omega_{A(n_A)}) \left\lbrace \mathbf{\Phi}^{u_L}\right\rbrace 
= \left\lbrace  \mathbf{0}  \right\rbrace \\
\end{array} \right\rbrace 
\label{wfesystem}
\end{equation}
Each line of system (\ref{wfesystem}) can be solved using for example Eq.\,(\ref{alternative_eigeq}), and provides $n$ eigensolutions $(\lambda_i,\mathbf{\Phi}_i)$:
\begin{equation}
 \mathbf{\Phi} = \left\lbrace 
 \begin{array}{c}
   \mathbf{\Phi}^{u_L}\\
   \mathbf{\Phi}^{u_R}
\end{array} \right\rbrace 
\end{equation}
Eigenvectors $\mathbf{\Phi}_i^{u_L}(\omega_{A(k)})$ stands for the deformed shape on the left side of a cell, associated with $i$th wave at $k$th wave cut-on frequency. Rectangular matrices A and B used for linear projection Eq.\,(\ref{matrices_A_B}) can be build upon these eigenvectors. 

\subsubsection{Properties of wave functions}
To summarize, the determination of a reduced set of eigenvectors is proposed to approximate full solutions over the whole frequency range, reducing the computation to one modal analysis and less than $n_A$ quadratic eigenproblems of size $n$, where $n_A$ is the number of cut-on frequencies in $\Delta \omega$. These vectors are independent of the cell length and a normalization can be performed to improve numerical conditioning.

\subsection{Wave basis optimization\label{3.3}}
In this section, the reduced wave basis for linear projection is determined using selected eigenvectors.
For the sake of clarity, unit-normalized deformed shape $\mathbf{\Phi}_i^{u_L}(\omega_{A(k)})$ will be written $\mathbf{\Phi}_i(k)$:
\begin{equation}
 \mathbf{\Phi}_i(k)=\mathbf{\Phi}_i^{u_L}(\omega_{A(k)})\quad , \quad {1\leq i\leq n}
\end{equation}

\subsubsection{Restriction to propagative waves}
At this point, no selection criteria was applied to the $n$ eigenvectors $\mathbf{\Phi}_i(k)$. If their wavenumbers are purely imaginary, they are evanescent. If a wavenumber is complex the wave is propagating and $|\Im(\kappa)|$ is the wave attenuation. In order to retain the least decaying waves and exclude highly dissipative waves, displacement can be expanded using only the positive-going propagative waves. This selection can be done for example retaining a number of eigenvectors for which $\Re(\kappa)\gg |\Im(\kappa)|$, say $\tilde{n}_k\ll n$ for the $k$th computed frequency.\\
The solutions subspace will be spanned using $\tilde{n}$ following vectors:
\begin{equation} 
\left\lbrace 
   \mathbf{\tilde{\Phi}}_i(k)\, , \, 1\leq i\leq \tilde{n}_k
\right\rbrace_{1\leq k\leq n_A} \quad  \textrm{and} \quad \tilde{n}=\sum_{k=1}^{n_A} \tilde{n}_k
\label{initialspan}
\end{equation}

\subsubsection{A simple optimization process for wave functions}
However, the concatenation of these wave shapes may involve redundant vectors, leading to a larger basis and numerical instabilities. This issue appears when two wave shapes $(\mathbf{\tilde{\Phi}}_i(k))_i$ and $(\mathbf{\tilde{\Phi}}_i(k+1))_i$ are computed at close frequencies. In this case, vectors associated to the same wave can be correlated. Denoting $\bar{.}$ and $.^T$ the complex conjugate and transposition, respectively, we can define correlation between two complex vectors using modal assurance criteria (MAC):
\begin{equation}
 \mathrm{Mac}(\mathbf{\phi},\mathbf{\psi})=\frac{(\mathbf{\phi}^T \mathbf{\bar{\psi}})\, (\mathbf{\psi}^T \mathbf{\bar{\phi}})}{(\mathbf{\phi}^T \mathbf{\bar{\phi}})\, (\mathbf{\psi}^T \mathbf{\bar{\psi}})}
\end{equation}
Let us define the orthogonality relation, in the sense of MAC correlation, between two vectors $\mathbf{\phi}$ and $\mathbf{\psi}$ as:
\begin{equation}
 \mathbf{\phi} \perp_\varepsilon \mathbf{\psi} \quad \Longleftrightarrow \quad \mathrm{Mac}(\mathbf{\phi},\mathbf{\psi})\leq \varepsilon
\end{equation}
where $\varepsilon \in \left[ 0\, , \, 1 \right] $ is a numerical criterion describing maximal tolerance for vector correlation in the reduced basis. We extend this definition to a vector's orthogonality to a set $\mathbf{\phi} \perp_\varepsilon \left\lbrace \mathbf{\psi}\right\rbrace$ , as the vector $\mathbf{\phi}$ being orthogonal to all the elements of $\left\lbrace \mathbf{\psi}\right\rbrace$. \\

In order to retain a limited number of sufficient vectors an iterative filtering procedure can be done, defining:
\begin{equation}
 \mathbf{\Psi}(1)=\left\lbrace \mathbf{\tilde{\Phi}}_i(1)\, , \, 1\leq i\leq \tilde{n}_1 \right\rbrace
 \label{psi1}
\end{equation}
then building the reduced basis from the recursive form:
\begin{equation}
\mathbf{\Psi}(k+1)=\left\lbrace \mathbf{\Psi}(k)\, , \, \mathbf{\tilde{\Psi}}(k+1) \right\rbrace
 \label{psi2}
\end{equation}
where $\mathbf{\tilde{\Psi}}(k+1)$ contains the vectors of $(\mathbf{\tilde{\Phi}}_i(k+1))_{1\leq i\leq \tilde{n}_{k+1}}$ which are orthogonal to the elements of $\left\lbrace \mathbf{\Psi}(1), ..., \mathbf{\Psi}(k)\right\rbrace $. The retained elements $\mathbf{\tilde{\Psi}}(k+1)$ are defined as:
\begin{equation}
 \mathbf{\tilde{\Psi}}(k+1) = \left\lbrace \chi \in (\mathbf{\tilde{\Phi}}_i(k+1))_{1\leq i\leq \tilde{n}_{k+1}} \quad \backslash \quad 
 \chi \perp_\varepsilon \mathbf{\Psi}(k) \right\rbrace 
 \label{psi3}
\end{equation}
The normalized and numbered vectors $(\mathbf{\tilde{\Psi}}_j(k))_{j,k}$ are denoted $\mathbf{\Gamma}_i$. It is also possible to evaluate wave correlation between two vectors $\mathbf{\Psi}_i(k)$ and $\mathbf{\Psi}_j(k+1)$ at frequencies $\omega_{A(k)}$ and $\omega_{A(k+1)}$ using Zhong and Williams \cite{Zhong1995} method, in which case the following product is maximized:
\begin{equation}
 (\mathbf{\Psi}_i(k))^T\left[ \mathbf{J}_n \right] (\mathbf{\Psi}_j(k+1))\quad 
 \textrm{where}\quad \left[ \mathbf{J}_n \right]=
 \left[ \begin{array}{cc}
         \mathbf{0} & \mathbf{I}_n\\
         -\mathbf{I}_n & \mathbf{0}
        \end{array}
\right]
\end{equation}

\subsubsection{Building the reduced basis}
Summarizing, the basis for wave expansion is defined using Eq.\,(\ref{eqappearance}) to compute a reduced set of frequencies used to form system Eq.\,(\ref{wfesystem}). Thus the reduced basis can be build from iterative procedure described Eqs.\,(\ref{psi1})--(\ref{psi3}). Considering hypothesis on numerical stability Eq.\,(\ref{stability}), all the vectors of matrix $A$ are real. This means that for an undamped waveguide vectors can be retained as they are defined earlier, whereas in dissipative cases it is preferable to display complex vectors $\Gamma_i$ as:
\begin{equation}
 \left[ \mathbf{\Gamma}^\ast_{i_1}\quad \mathbf{\Gamma}^\ast_{i_2}\right] = \left[ \Re(\mathbf{\Gamma}_i) \quad \Im(\mathbf{\Gamma}_i)\right] 
 \label{notationcplx}
\end{equation}
where $\Re$ and $\Im$ denote real and imaginary parts. However, the use of complex vectors for the reduced basis theoretically leads to the same conclusions, providing hermitian product for matrices projection. In most cases, wave dispersion is required on a frequency range starting at 0\,Hz. Then, a set of undeformed motions, computed using system Eq.\,(\ref{rigidmotion}), is enough to describe first-order shape functions.
\begin{equation}
\left[ \begin{array}{cc}
 \mathbf{K}_{LL} & \mathbf{K}_{LR} \\
 \mathbf{K}_{RL} & \mathbf{K}_{RR}
\end{array}\right] 
\left\lbrace \begin{array}{c}
 \mathbf{q}_{L} \\
 \mathbf{q}_{R}
\end{array}\right\rbrace 
= \left\lbrace \begin{array}{c}
 \mathbf{0} \\
 \mathbf{0}
\end{array}\right\rbrace 
\label{rigidmotion}
\end{equation}
These real vectors, say $n_c$, represent cell rigid motions of the waveguide under first-order waves.

\subsection{Formulation of the reduced eigenproblem}
\subsubsection{Projection of dynamic stiffnesses}
Considering sections (\ref{3.1}) and (\ref{3.3}), matrix $\mathcal{B}_{(n,r)}$ defined Eq.\,(\ref{matrices_A_B}) can be expressed using notations Eq.\,(\ref{notationcplx}) as:
\begin{equation}
\mathcal{B} = \left[\mathbf{\Gamma}_1 \, ,...,\,\mathbf{\Gamma}_{n_c} \, , \mathbf{\Gamma}^\ast_{n_c+1} \, ,...,\, \mathbf{\Gamma}^\ast_{n_c+r} \right] 
\label{defineB}
\end{equation}
where $r$ is the sum of $n_c$ and the number of retained wave functions selected among the $\tilde{n}$ vectors defined Eq.\,(\ref{initialspan}). Spectral problem Eq. (\ref{spectral_system}) can now be rewritten:
\begin{equation}
\left[ \lambda \mathbf{\tilde{D}}_{LR}(\omega)+(\mathbf{\tilde{D}}_{LL}(\omega)+\mathbf{\tilde{D}}_{RR}(\omega))+\frac{1}{\lambda}\mathbf{\tilde{D}}_{RL}(\omega)\right] \boldsymbol{\mu}_L=\mathbf{0}
\end{equation}
where $\mathbf{\tilde{D}}_{ij}$ stands for the projected dynamic stiffnesses:
\begin{equation}
\mathbf{\tilde{D}}_{ij}=\mathcal{B}^T \mathbf{D}_{ij} \mathcal{B}
\end{equation}
The $r\times r$ eigenproblem can now be written:
\begin{equation}
 \left[\begin{array}{cc}
  -\mathbf{\tilde{D}}_{RL} & -(\mathbf{\tilde{D}}_{LL}+\mathbf{\tilde{D}}_{RR}) \\
  \mathbf{0} & -\mathbf{\tilde{D}}_{RL}
 \end{array}\right]
 \left\lbrace \begin{array}{c}
  \boldsymbol{\mu}_L \\
  \boldsymbol{\mu}_R
 \end{array}\right\rbrace 
 = \lambda \left[\begin{array}{cc}
  \mathbf{0} & \mathbf{\tilde{D}}_{LR} \\
  -\mathbf{\tilde{D}}_{RL} & \mathbf{0}
 \end{array}\right] 
 \left\lbrace \begin{array}{c}
  \boldsymbol{\mu}_L \\
  \boldsymbol{\mu}_R
 \end{array}\right\rbrace 
 \label{reducedeig}
\end{equation}

\subsubsection{Desired eigensolutions and error evaluation}
The reduced eigenproblem can also be formulated:
\begin{equation}
 \mathbf{\tilde{S}}\mathbf{\tilde{\Phi}}=\tilde{\lambda}\mathbf{\tilde{\Phi}}
\end{equation}
where 
\begin{equation}
 \mathbf{\tilde{S}}=
 \left[ \begin{array}{cc}
  \mathbf{0} & \mathbf{I} \\
  \mathbf{\tilde{D}}_{LR}^{-1}\mathbf{\tilde{D}}_{RL} & \mathbf{\tilde{D}}_{LR}^{-1}(\mathbf{\tilde{D}}_{LL}+\mathbf{\tilde{D}}_{RR})
 \end{array}\right] 
\end{equation}
and the $(r\times1)$ reduced eigenvectors are denoted:
\begin{equation}
\mathbf{\tilde{\Phi}} =  \left\lbrace \begin{array}{c}
  \boldsymbol{\mu}_L \\
  \boldsymbol{\mu}_R
 \end{array}\right\rbrace 
\end{equation}
Reduced eigenvectors are written in spatial coordinates $\mathbf{\bar{\Phi}} = \mathfrak{B} \mathbf{\tilde{\Phi}}$, as:
\begin{equation}
 \mathfrak{B} = \left[\begin{array}{cc}  
 \mathcal{B} & \mathbf{0} \\
 \mathbf{0} & \mathcal{B} \\
\end{array}\right]
\label{base_B}
\end{equation}
The expanded $(n\times n)$ transfer matrix is defined as well: $\mathbf{\bar{S}} = \mathfrak{B} \mathbf{\tilde{S}} \mathfrak{B}^T$. The error on the reduced system eigensolutions $(\tilde{\lambda} , \tilde{\mathbf{\Phi}})$, induced by the aforementioned basis reduction can be expressed as:
\begin{equation}
 \varepsilon = \dfrac{\|(\mathbf{S} - \overline{\mathbf{S}})\bar{\mathbf{\Phi}}\|}{\|\bar{\mathbf{\Phi}}\|}
\end{equation}
As $\mathbf{\bar{S}}\mathbf{\bar{\Phi}} = \tilde{\lambda}\mathbf{\bar{\Phi}}$, we obtain:
\begin{equation}
 \varepsilon = \dfrac{\|(\mathbf{S} - \tilde{\lambda}\mathbf{I})\bar{\mathbf{\Phi}}\|}{\|\bar{\mathbf{\Phi}}\|}
 \label{eq:error}
\end{equation}
Then, the error induced by the wave base truncation can be defined by injecting the reduced eigensolutions in the full eigenproblem. Thus no knowledge of the exact solutions $(\lambda , \mathbf{\Phi})$ is required.

\subsubsection{Discussion}
The size, $r$, of the reduced eigenproblem described above depends on the number of propagating waves in the frequency band $\Delta\omega$. The number of resolutions, $n_A$, of the full eigenproblem is independent of the number of frequency samples $N_{fs}=\nicefrac{\Delta\omega}{\delta\omega}+1$. As a result, the computational cost will be greatly reduced, solving $n_A \ll N_{fs}$ full eigenproblems of size $(n \times n)$ and $N_{fs}$ reduced problems of size $(r \times r) \ll (n \times n)$.

\subsection{Algorithm for the reduced WFE method}
\begin{figure}[!ht]
  \centering
   \includegraphics[width=0.8\columnwidth]{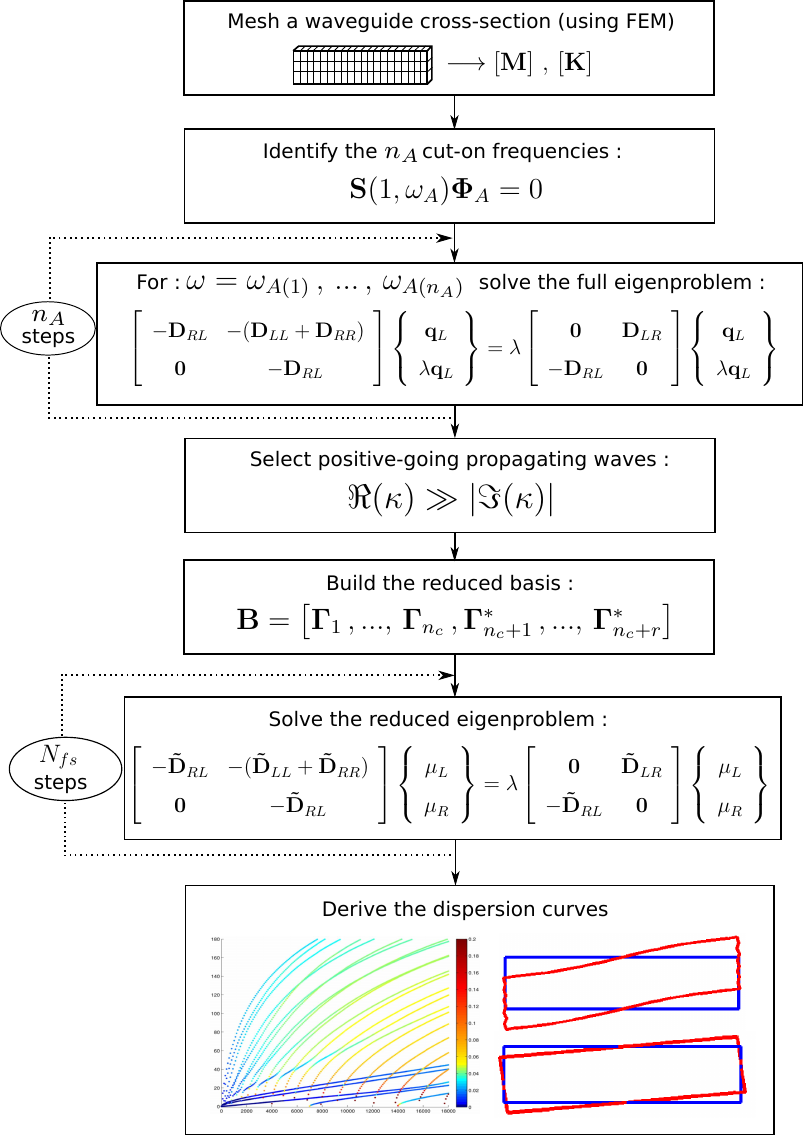}
   \caption{Procedure for the reduced wave finite element method.}
   \label{algo}
\end{figure}

The proposed reduction strategy is illustrated Fig. \ref{algo}, each step is detailed as follows:
\begin{itemize}[label=$\triangleright$]
 \item Build the mass and stiffness matrices of the waveguide cross-section and reorder as described Eq.\,(\ref{eqmotion}).
 \item Identify the cut-on frequencies over the frequency band using Eq.\,(\ref{eqappearance}).
 \item Solve the $n_A$ eigenproblems defined Eq.\,(\ref{wfesystem}).
 \item Select eigenvectors associated with propagating waves using Section \ref{3.3}.
 \item Build the wave basis Eq.\,(\ref{defineB}) using the optimization process proposed Eqs.\,(\ref{psi1})--(\ref{psi3}).
 \item Solve the $N_{fs}$ reduced eigenproblems of size $(r \times r)$ given Eq.\,(\ref{reducedeig}).
 \item Derive eigensolutions using projection defined Eq.\,(\ref{base_B}).
\end{itemize}

\section{Numerical applications\label{numerical}}
\subsection{Example 1: Sandwich beam}
\begin{figure}[!ht]
  \centering
   \includegraphics[width=0.5\columnwidth]{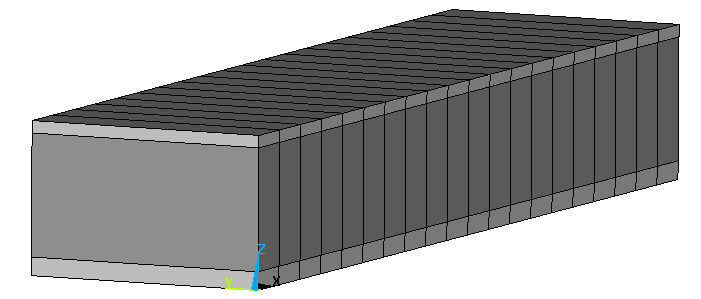}
   \caption{Illustration: sandwich beam.}
   \label{sandwich_beam}
\end{figure}

In this section, three solutions for free wave propagation in the sandwich beam illustrated Fig. \ref{sandwich_beam}, are compared. An analytical formulation depicted in Ref.\,\cite{Banerjee2005}, and results obtained from WFE method in Ref.\,\cite{Mencik2008} are compared to a WFE formulation using a refined FEM for the cross-sectional discretization. The limitations of a coarse mesh are discussed and the requirement for a refined mesh for the WFE computation is highlighted. Then, the proposed reduction strategy is applied to the refined FEM model and results are discussed.

\subsubsection{Description}
\begin{figure}[!ht]
  \centering
   \includegraphics[width=0.4\columnwidth]{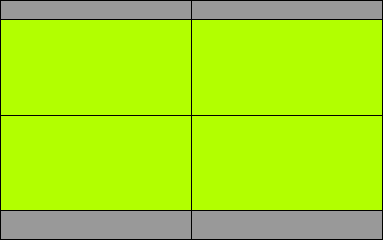}
   \includegraphics[width=0.4\columnwidth]{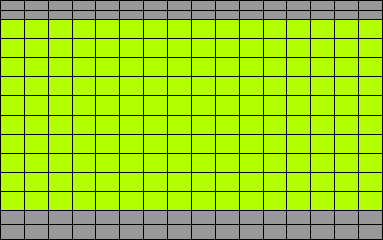}
   \caption{Cross-section discretization using a coarse mesh (left) and a refined mesh (right).}
   \label{mesh_sandwich}
\end{figure}

The rectangular cross-section three-layered sandwich beam is composed of a rubber core surrounded by two steel layers. Detailed description of the layers is given Table \ref{table:beam}. The width of the beam is 40\,mm and the total thickness is 25\,mm. A constant loss factor $\eta = 0.01$ is assumed for the numerical modeling while no damping is considered in the analytical formulation.
\begin{table}[!ht]
\begin{center}
\begin{tabular}{|l|l|r|r|r|}
  \hline
  Layer -- & Thickness & Density  & Young Modulus  & Poisson  \\
  material & (mm) & (kg\,m$^{-3}$) & (Pa) & coefficient \\
  \hline
  3 -- Steel skin & $2$ & $7850$ & $210\times10^9$ & $0.3$ \\
  \hline
  2 -- Rubber core & $20$ & $950$ & $1.5\times10^6$ & $0.48$ \\
  \hline
  1 -- Steel skin & $3$ & $7850$ & $210\times10^9$ & $0.3$ \\
  \hline
\end{tabular} 
\caption{Material properties for the sandwich beam layers.\label{table:beam}}
\end{center}
\end{table}
The coarse mesh used in Ref.\,\cite{Mencik2008} has 8 linear block elements having 3 degrees of freedom (DOF) per node and a total of 45 DOFs. The proposed refined mesh has 224 linear elements and 1530 DOFs, both cells have a length $d=2\times10^{-3}$\,m in the direction of propagation, their meshes are shown Fig. \ref{mesh_sandwich}.

\subsubsection{Refined FEM model for the WFE method}
\begin{figure}[!ht]
  \centering
   \includegraphics[width=0.8\columnwidth]{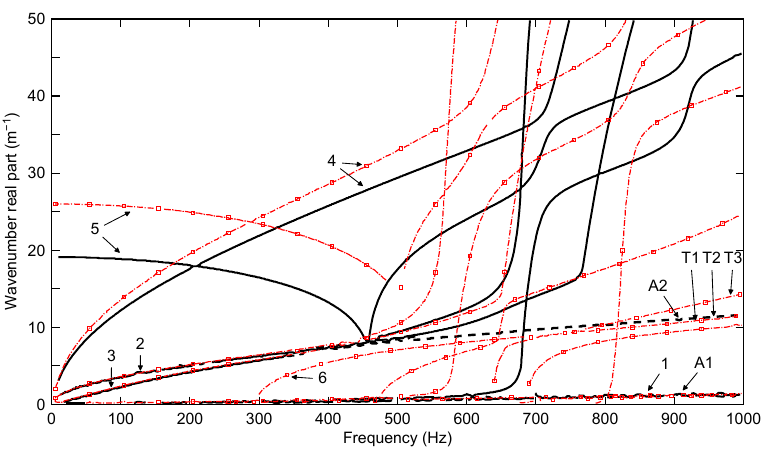}
   \caption{Dispersion curves for the global waveguide. Refined WFE: (red -$\,${\tiny$\square$}$\,$-$\,${\tiny$\square$}$\,$-), Coarse WFE: (black ------), Analytical solution: (black - - - -).}
   \label{coarse_VS_refined}
\end{figure}

Real wavenumbers obtained using the analytical, coarsely meshed WFE and finely meshed WFE formulations are presented Fig. \ref{coarse_VS_refined} over the frequency range $\Delta \omega = [0 , 1000\,\mathrm{Hz}]$.
Below 450\,Hz: coarse and refined WFE formulations provide similar results for the longitudinal (1), torsional (2) and transverse flexural (3) waves. The flexural (4) wave slightly differs from the coarse model, while the complex thickness contractional (5) is clearly different between the two models. Note that the cut-on frequency for this last wave is 500\,Hz, instead of 470\,Hz for the coarse model, and a second-order longitudinal wave (6) cuts-on at 300\,Hz in the refined model only.\\

\begin{figure}[!ht]
  \centering
   \includegraphics[width=0.8\columnwidth]{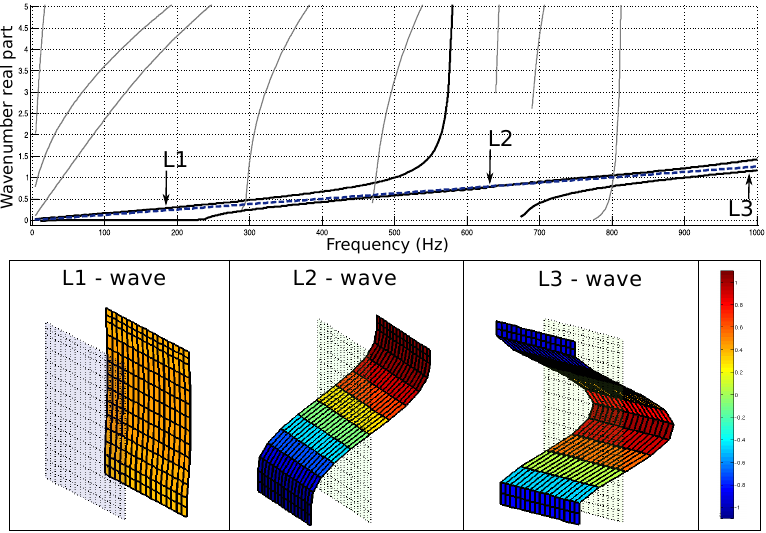}
   \caption{Deformed shapes associated with 3 orders of longitudinal waves and detailed wavenumbers.}
   \label{longit_waves}
\end{figure}

\begin{figure}[!ht]
  \centering
   \includegraphics[height=8cm]{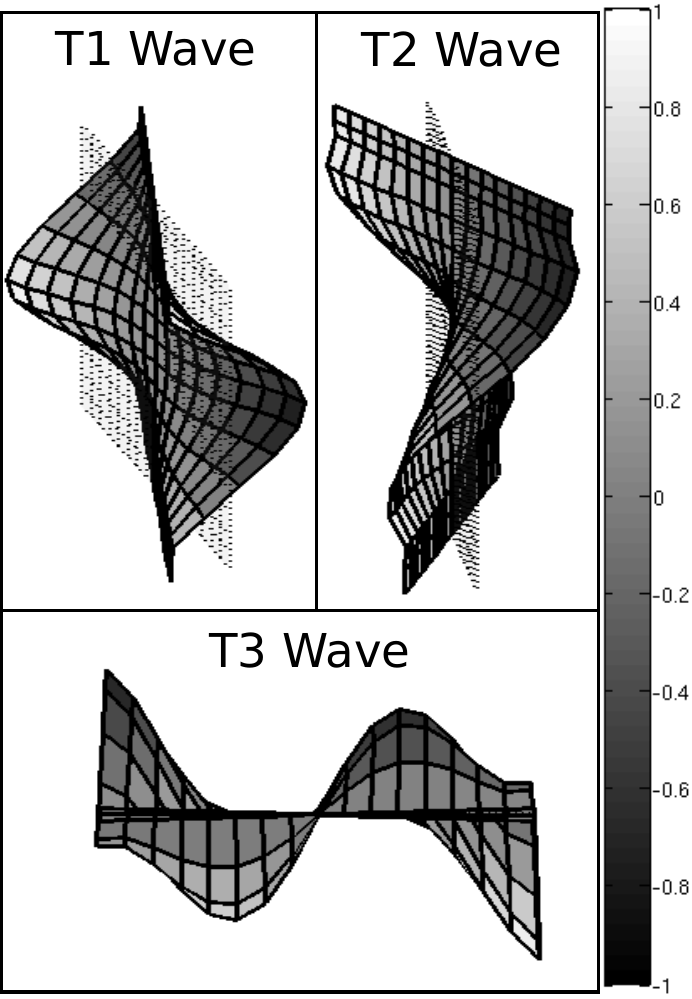}
   \caption{Deformed shapes associated with three transverse waves. T1 and T2: 3D view.}
   \label{transverse_1_2_3}
\end{figure}

After 500\,Hz: the analytical solution (A1) and both WFE formulations appear to be consistent for the longitudinal wave (1). Fig. \ref{longit_waves} shows a zoom of the propagating wavenumbers previously depicted in Fig. \ref{coarse_VS_refined}. Three orders of longitudinal waves are highlighted, compared to the analytical solution and illustrated. Only the refined model differentiates the 3 longitudinal wave modes described Fig. \ref{longit_waves}. It can also be seen that several forms of longitudinal waves can coexist at the same frequencies. L2 and L3 waves exhibit the impedance mismatch between the skins and the soft core of the beam. The coarse model accuracy for these waves can be explained by the possibility to approximate them using 4 linear elements in the beam thickness.
The coarse model fails to compute the transverse flexural wave (A2) predicted by the analytical formulation. This wave mode appears to require a finer meshing for the inner rubber core. However, the analytical solution, based on a thickness homogenization does not entirely describe the waveguide behavior. Indeed, between 800\,Hz and 1000\,Hz three localized transverse flexural waves, presented Fig. \ref{transverse_1_2_3}, are propagating. They are associated to the transverse wave localization in the skins or in the core. T1 is the symmetric transverse wave with opposed skin-core phase, T2 is the antisymmetric transverse wave localized in the skins, T3 is the local third-order transverse wave in the core.
Most of the wave modes from the coarse model are incorrect after 500\,Hz. At 750\,Hz, artificial veering occurs around $k=15$ when the wave shapes turns more warped, leading to possible confusions in the wave modes matching.\\

The importance of a finely meshed model for the application of WFE method to a sandwich beam has been confirmed. The coarse model presented above fails to predict the wave propagation in this waveguide, especially over 500\,Hz, and most of the wave shapes computed are erroneous. However, accurate dispersion curves and wave shapes can be obtained using the WFE method, under the condition that the model is highly discretized and the sampling frequency is small enough to match properly the wave modes.

\subsubsection{Reduced WFE formulation}
\begin{figure}[!ht]
  \centering
   \includegraphics[width=0.75\columnwidth]{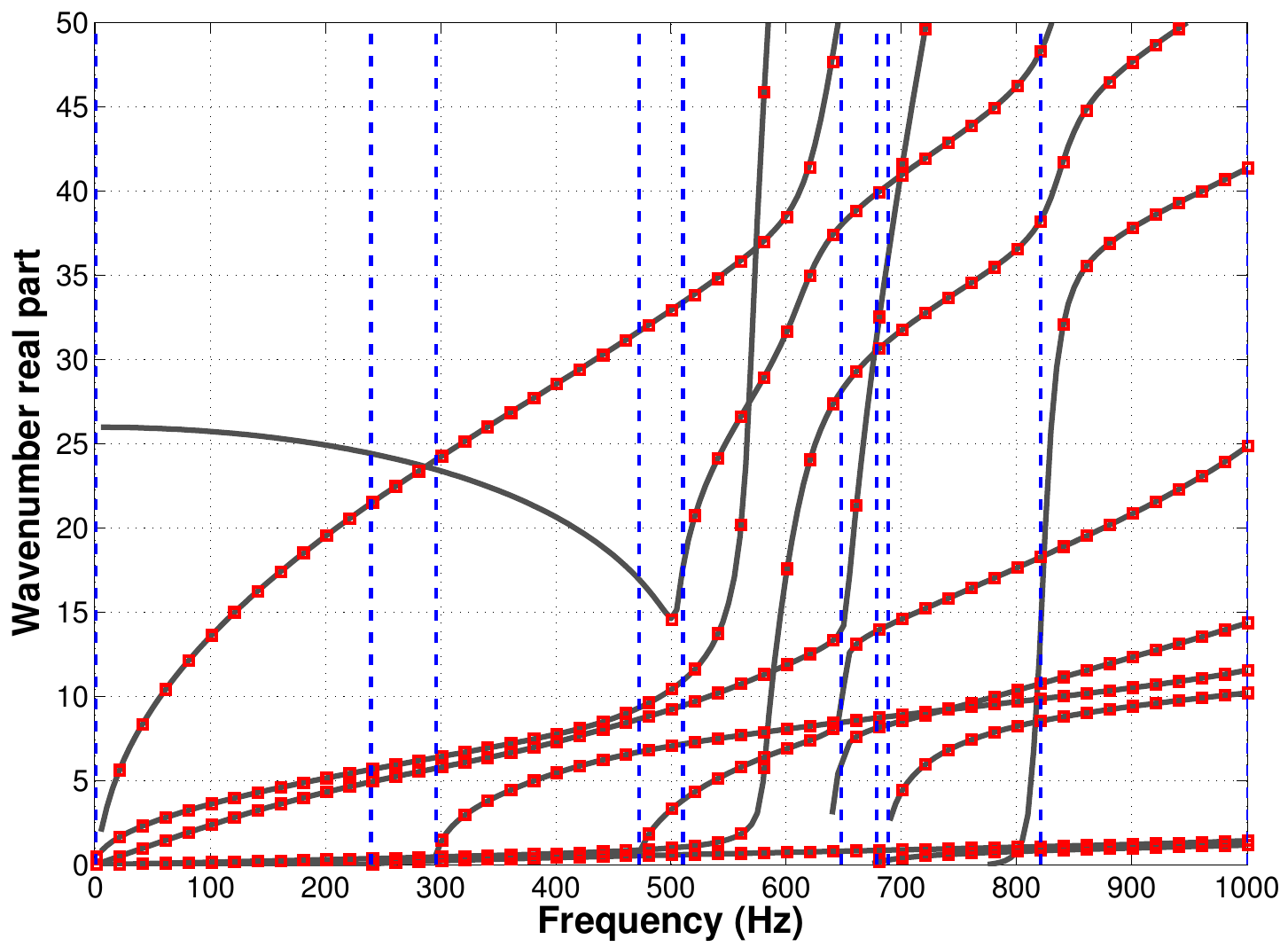}
   \caption{Comparison between reduced (red {\tiny$\square$}) and full (black ---) wavenumbers for the sandwich waveguide. Vertical blue lines stand for predicted cut-on frequencies.}
   \label{reduced}
\end{figure}

\begin{figure}[!ht]
  \centering
   \includegraphics[width=0.7\columnwidth]{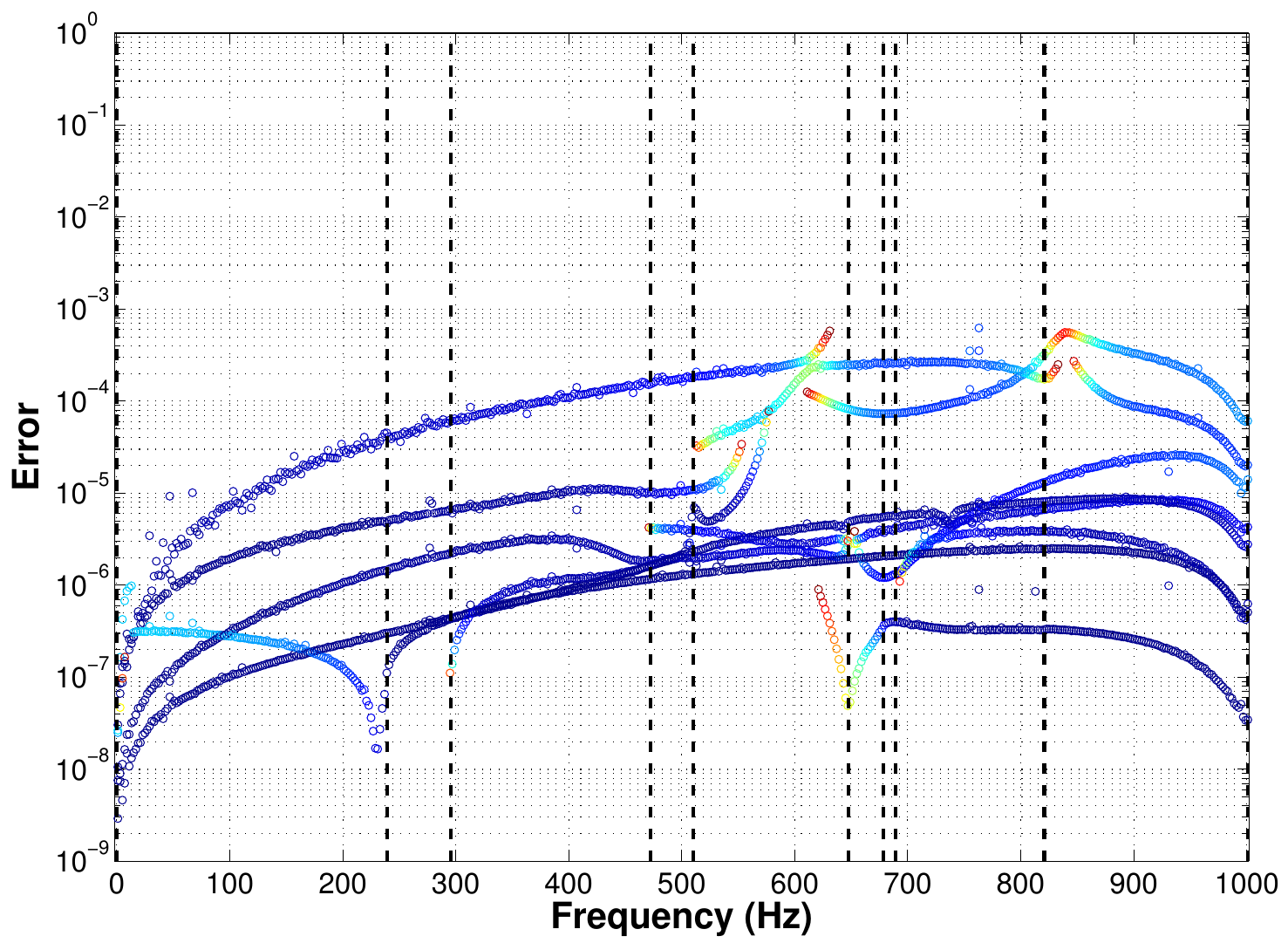}
   \caption{Error $\varepsilon$ for the propagating waves eigenvectors ($|Im(k)| < 1$).}
   \label{fig:error}
\end{figure}

The proposed reduction strategy is applied to the refined model described above and compared to the classical WFE formulation. Real wavenumbers are presented Fig. \ref{reduced}. Vertical lines denote the cut-on frequencies, where the the full eigenproblems are solved. Note that all of the propagating waves starts at the predicted cut-on frequencies. Other cuts-on are linked to important veerings associated to a propagating wave appearance, such as the thickness contractional wave at 500\,Hz, or complex waves turning into propagating waves. There are 4 primary waves at 0\,Hz and 8 cut-on frequencies, which is consistent with the 12 final branches identified on the dispersion curves Fig. \ref{reduced}. This result shows that the number of propagating waves was correctly predicted using Eq.\,(\ref{eqappearance}).
Very good correlation is observed between the classical formulation and the reduced model, built only using 10 eigensolutions: at the 2 frequency range boundaries and the 8 cut-on frequencies. The reduced wave basis is computed using a correlation criteria $\varepsilon = 0.6$, and contains 124 vectors instead of 1530 for the full system. Veering phenomena, where important variations of the wave shapes can occur, are accurately estimated using the reduced wave basis, see Fig. \ref{reduced} at 650\,Hz for the wavenumber $k \simeq 13$. The error on the wave shapes, defined Eq.\,(\ref{eq:error}) is presented Fig. \ref{fig:error} for the main propagating waves. Wave matching is provided in order to compare the errors associated with each propagating wave. A color scatter is also proposed to identify mostly propagating wave modes. It shows a good accuracy between full and reduced WFE, the maximal error being around $5\times10^{-4}$. Cut-on frequencies are identified by vertical dashed lines. At 230\,Hz, 300\,Hz and 650\,Hz, it is noticeable that each appearing wave shapes retained provide accurate solutions around the cut-on frequency.
Significant reduction of CPU time is observed for the proposed WFE formulation, a detailed description is given Table \ref{table:CPU}. This method provides a $97\%$ reduction of CPU time in comparison with classical WFE formulation, and provides accurate dispersion curves and wave shapes. Note that most of the computation time for the reduced model is due to the 10 full computations, and similar accuracy was obtained by increasing the correlation criteria $\varepsilon = 0.95$ and using only 6 full computations, leading to a 148 vectors basis.
\begin{table}[!ht]
\begin{center}
\begin{tabular}{|l|c|c|r|}
  \hline
  \textbf{Classical WFE} & Number of & Time per & CPU time \\
   & iterations & iteration & \\
  \hline
  Resolution & & & \bf{Total:}\\
  $[0 , 1000\,\mathrm{Hz}]$ per $2\,\mathrm{Hz}$ & $\times$500 & 152\,s & \bf{21\,h 07\,m} \\
  \hline
  \hline
  \textbf{Reduced WFE} & Number of & Time per & CPU time \\
   & iterations & iteration & \\
  \hline
  Appearance $\omega_{A(k)}$ & $\times$1 & 16\,s & 16\,s \\
  \hline
  Full solution at $\omega_{A(k)}$ & $\times$10 & 152\,s & 25\,m 20\,s\\
  \hline
  Building reduced basis  & $\times$1 & 132\,s & 2\,m 12\,s \\
  \hline
  Resolution & & & \\
  $[0 , 1000\,\mathrm{Hz}]$ per $2\,\mathrm{Hz}$ & $\times$500 & 0.09\,s & 45\,s \\
  \hline
   & & & \bf{Total: 28\,m 33\,s}\\
  \hline
\end{tabular}
\caption{Sandwich beam: CPU time for the full and reduced WFE formulations.\label{table:CPU}}
\end{center}
\end{table}

\subsubsection{Conclusions}
In this example, the proposed reduced WFE formulation was applied to a finely discretized cross-section of a sandwich beam. First, the requirement for a refined FEM model was demonstrated by comparing the WFE solutions with an analytical solution and a coarsely meshed WFE model. Higher orders for propagating waves were identified. Then, the proposed reduced formulation was compared to a full computation and showed a very good agreement for both wavenumbers and wave shapes. Note that the proposed criteria for the choice of full computation frequencies can be excessive when numerous waves are propagating and lead to unnecessary rich wave basis lowering the numerical reduction efficiency.

\subsection{Example 2: Elasto-acoustic waveguide}
\subsubsection{Description}
\begin{figure}[!ht]
  \centering
   \includegraphics[width=0.69\columnwidth]{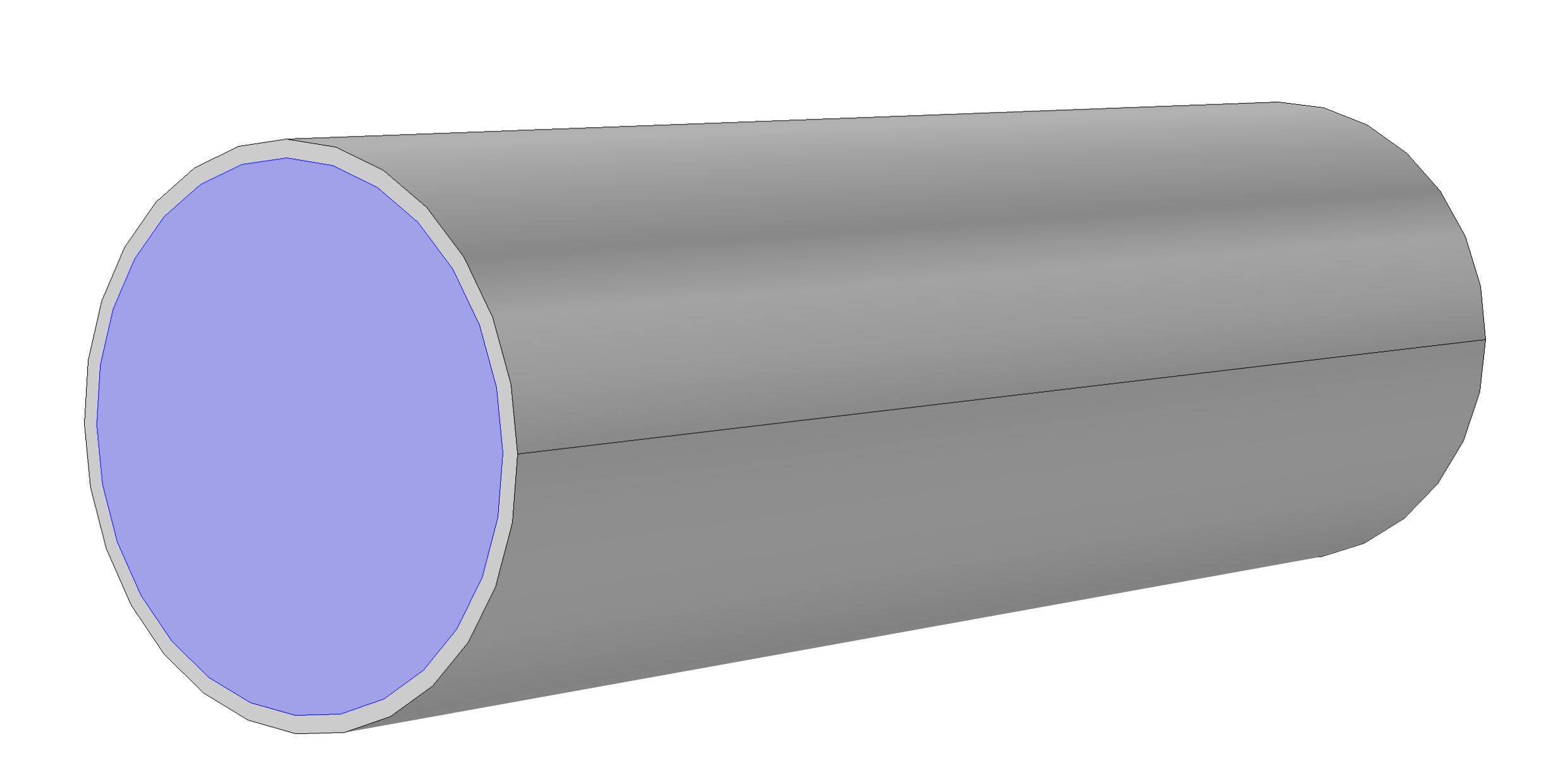}
   \includegraphics[width=0.3\columnwidth]{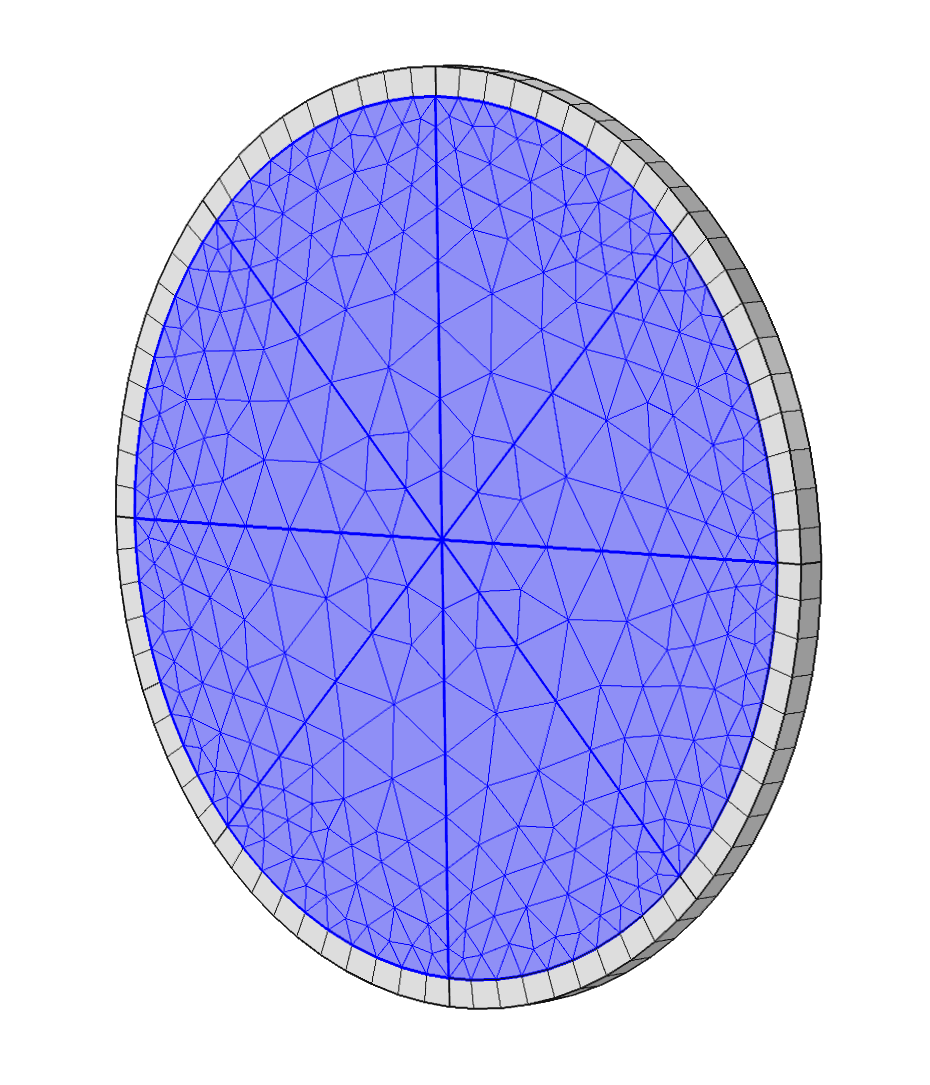}
   \caption{(a) Illustration of an elastic pipe filled with acoustic fluid. (b) Finite element model of a representative section of the elasto-acoustic pipe.}
   \label{pipe}
\end{figure}

As mentioned earlier, guided waves have numerous applications in structural dynamics. Elasto-acoustic waveguides are also a topic of interest, especially in SHM for crack and leakages detection. The theory of the propagation of vibro-acoustic waves in fluid-filled, cylindrical elastic shells has been studied in detail by Fuller and Fahy 1982 \cite{Fuller1982}. This example provides an extension of the reduced WFE formulation for a coupled fluid--structure system, where the inner fluid involves an important number of degrees of freedom to ensure decent discretization with FEM. The WFE formulation has been proposed by Mencik and Ichchou \cite{Mencik2007} to predict the forced response of a fluid-filled pipe shown Fig. \ref{pipe}a. 
The reduced WFE formulation can be applied as detailed Section \ref{wfe_classique} by using the dynamic stiffnesses described in \cite{Mencik2007}, which exhibit quadratic dependence with respect to $\omega$. A formulation is proposed in \ref{annexe2} to determine the cut-on frequencies by solving a linear eigenproblem in $\omega^2$ instead of a quadratic problem in $\omega$. In this example we consider a similar elastic pipe filled with a homogeneous, barotropic and compressible fluid whose viscosity is neglected. The cross-section shown Fig. \ref{pipe}b is defined as $R^e-R^i=h$ with $R^e=80\,\mathrm{mm}$ and $h=5\,\mathrm{mm}$, the pipe has a Young modulus $E^s=190\,\mathrm{Gpa}$, density $\rho^s=7800\,\mathrm{kg\,m^{-3}}$ and Poisson ratio $\nu=0.29$. The fluid has a sound velocity $c^a=1484\,\mathrm{m\,s^{-1}}$ and a density $\rho^a=1000\,\mathrm{kg\,m^{-3}}$. The finite element model of a representative section of length $d=5\,\mathrm{mm}$ involves three-dimensional linear elements and $n=n^s+n^a=1706$ degrees of freedom where superscripts $s$ and $a$ refer to solid and fluid domains. 

\subsubsection{Results}
\begin{figure}[!ht]
  \centering
   \includegraphics[width=0.8\columnwidth]{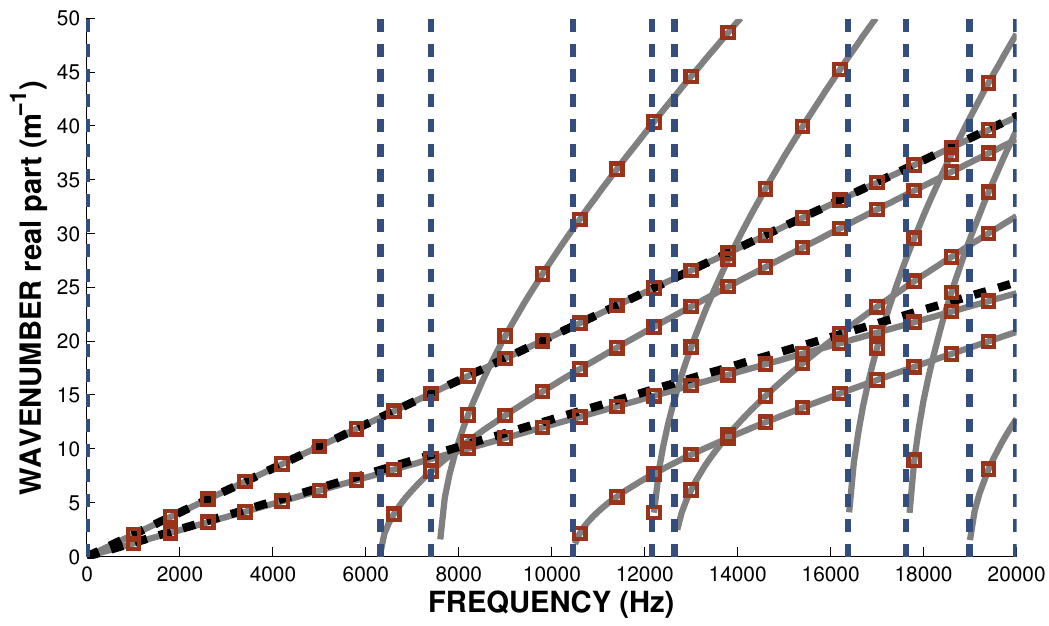}
   \caption{Comparison between reduced (red {\tiny$\square$}) and full (black ---) WFE formulations for the elasto-acoustic waveguide. Analytical low frequency solutions for axial quasi-longitudinal extensional and torsional waves (- - -).}
   \label{pipe_realK}
\end{figure}

\begin{figure}[!ht]
  \centering
   \includegraphics[width=0.75\columnwidth]{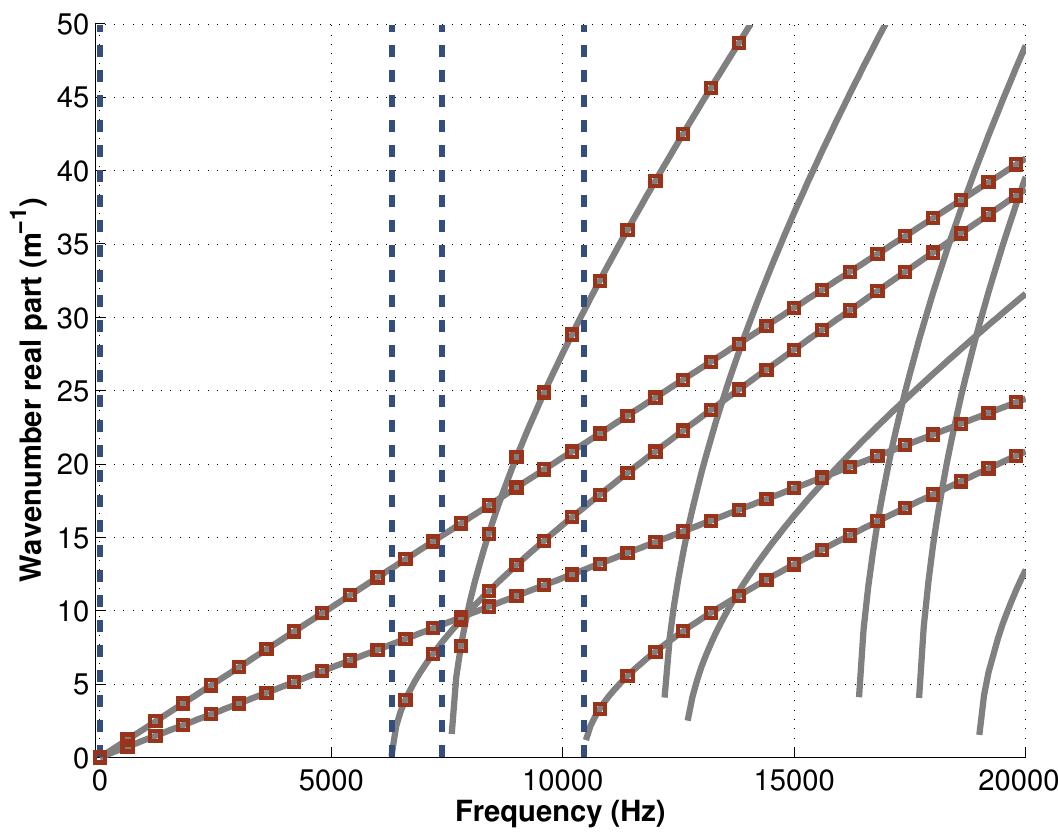}
   \caption{Wavenumbers for an elasto-acoustic waveguide. Reduced WFE formulation: (red {\tiny$\square$}). Full WFE (gray ---). }
   \label{pipe_partial}
\end{figure}

The dispersion curves are computed at frequencies between $0$\,Hz and $20$\,kHz and wavenumbers real parts are presented Fig. \ref{pipe_realK}. 
A very good correlation is observed between the two WFE formulations over the whole frequency band. The 8 cut-on frequencies are identified and 10 full computations are performed. The analytical solution is given for torsional and longitudinal waves at low frequencies and is consistent with numerical solutions. The reduced wave basis is computed using a correlation criteria $\varepsilon = 0.8$ and contains 148 vectors. Significant reduction of CPU time is observed with this reduced formulation, a detailed description is given Table \ref{table:CPU2}. The reduction provides a $94\%$ reduction of CPU time in comparison with classical WFE formulation, and gives accurate dispersion curves and wave shapes.
\begin{table}[!ht]
\begin{center}
\begin{tabular}{|l|c|c|r|}
  \hline
  \textbf{Classical WFE} & Number of & Time per & CPU time \\
   & iterations & iteration & \\
  \hline
  Resolution & & & \bf{Total:} \\
  $[0 , 20{,}000\,\mathrm{Hz}]$ per $100\,\mathrm{Hz}$ & $\times$200 & 215\,s & \bf{11\,h 56\,m 40\,s} \\
  \hline
  \hline
  \textbf{Reduced WFE} & Number of & Time per & CPU time \\
   & iterations & iteration & \\
  \hline
  Appearance $\omega_{A(k)}$ & $\times$1 & 28\,s & 28\,s \\
  \hline
  Full solution at $\omega_{A(k)}$ & $\times$10 & 215\,s & 2150\,s \\
  \hline
  Building reduced basis  & $\times$1 & 196\,s & 196\,s \\
  \hline
  Resolution & & & \\
  $[0 , 20{,}000\,\mathrm{Hz}]$ per $100\,\mathrm{Hz}$ & $\times$200 & 0.08 & 16\,s \\
  \hline
   & & & \bf{Total: 39\,m 50\,s}\\
  \hline
\end{tabular}
\caption{Pipe filled with fluid: CPU time for the full and reduced WFE formulations.\label{table:CPU2}}
\end{center}
\end{table}
The accuracy of this method relies on the number and frequencies of the full computations retained. A reduced basis containing only low frequency eigenvectors is expected to produce significant errors as the frequeny induces changes of the wave shapes. An example is given Fig. \ref{pipe_partial}, where the wavenumbers are computed from 0\,Hz to 20\,kHz, while only 4 cut-on frequencies are retained. The correlation criteria is $\varepsilon = 0.8$ and propagating waves are retained if $|\Im(k)| \leq 0.5$, leading to a $(60\times60)$ reduced system. A good accuracy is shown for the waves whose cut-on frequencies are retained. Wavenumbers are correct until 20\,kHz while the last eigenvector is computed at 10.5\,kHz. This result indicates that the reduced system solutions are consistent with the number of wave shapes retained for the reduced basis. However, no solutions are found for the higher-order waves. Note that the number of eigenvectors needed for the reduced basis can be overestimated if an important number of complex waves are retained. This issue can occur in highly dissipative structures.

\subsubsection{Conclusions}
In this example, the proposed reduced WFE formulation is applied to a cylindrical elastic shell filled with fluid. Dispersion curves are accurately determined with a reduction of $94\%$ of the computation time. As expected, a truncated wave basis  may not accurately predict the system solutions, the errors are consistent with the frequencies retained for the reduced wave base computations.

\section{Concluding remarks}
The main limitation of the WFE method for the broadband analysis of composite waveguides is due to their cross-sectional discretization. Coarsely meshed model produces inaccurate solutions and restrains the analysis to low-order propagating phenomena, especially within  the field of composite structures where internal components can enable numerous localized wave modes. In this paper a reduction strategy was presented to study wave propagation in such waveguides. It provides a significant reduction of the computation time and enables broadband WFE analysis for a wide range of composite waveguides with complex cross-sectional geometries. The method was successfully applied to a refined model of a laminated composite beam. An error evaluation on the reduced eigenvectors was proposed and high-order localized wave modes were accurately predicted although a $97\%$ reduction of the computation time. An elasto-acoustic pipe was considered. A adapted formulation was provided to identify the cut-on frequencies from a linear eigenproblem. The influence of a truncated basis was evaluated and showed good agreement with the proposed wave selection. Furthermore, the proposed method might be used to study high-order waves in composite waveguides such as wave localization and couplings. Further applications will be reported.

\section*{Acknowledgement}
The authors would like to gratefully acknowledge Airbus Helicopters for their financial support.

\newpage
\appendix

\section*{Appendix A. An extension for the fluid--structure interaction}
\refstepcounter{section}\label{annexe2}
\setcounter{equation}{0}
\renewcommand{\theequation}{A.\arabic{equation}}
Considering the (U, p) formulation for the elasto-acoustic problem \cite{Mencik2007}:
\begin{equation}
-\omega^2 \left[ \begin{array}{cc}
\mathbf{M}^s & \mathbf{0} \\
\rho_0\mathbf{C}^T & \mathbf{M}^a
\end{array} \right] 
\left\lbrace  \begin{array}{c}
   \mathbf{u} \\
   \mathbf{p}
\end{array} \right\rbrace +
\left[ \begin{array}{cc}
\mathbf{K}^s & -\mathbf{C} \\
\mathbf{0} & \mathbf{K}^a
\end{array} \right] 
\left\lbrace  \begin{array}{c}
   \mathbf{u} \\
   \mathbf{p}
\end{array} \right\rbrace =
\left\lbrace  \begin{array}{c}
   \mathbf{F}^s\\
   \mathbf{F}^a
\end{array} \right\rbrace
\label{elasto_eq}
\end{equation}
where $\mathbf{M}^s$, $\mathbf{K}^s$, $\mathbf{M}^a$, $\mathbf{K}^a$ stand for the structural and acoustic masses and stiffnesses, $\mathbf{C}$ is the fluid--structure coupling matrix, and $\mathbf{F}^s$, $\mathbf{F}^a$ are the structural and acoustic forces vectors. Following the steps described Section~\ref{wfe_reduite} and using same notation and $D^{s,a} = K^{s,a}-\omega^2 M^{s,a}$, the spectral problem can be written:
\begin{multline}
\left[ \begin{array}{cc}
\lambda \mathbf{D}_{LR}^s + \frac{1}{\lambda} \mathbf{D}_{RL}^s + \mathbf{D}_{LL}^s + \mathbf{D}_{RR}^s &
 - \left(  \lambda \mathbf{C}_{LR} + \frac{1}{\lambda} \mathbf{C}_{RL} + \mathbf{C}_{LL} + \mathbf{C}_{RR} \right)  \\
-\rho_0 \omega^2 \left( \lambda \mathbf{C}_{LR}^T + \frac{1}{\lambda} \mathbf{C}_{RL}^T + \mathbf{C}_{LL}^T + \mathbf{C}_{RR}^T\right) & 
\lambda \mathbf{D}_{LR}^a + \frac{1}{\lambda} \mathbf{D}_{RL}^a + \mathbf{D}_{LL}^a + \mathbf{D}_{RR}^a
\end{array} \right] \\
\times \left\lbrace  \begin{array}{c}
   \mathbf{u}\\
   \mathbf{p}
\end{array} \right\rbrace = 
\left\lbrace  \begin{array}{c}
   \mathbf{0}\\
   \mathbf{0}
\end{array} \right\rbrace
\label{spectral_elasto}
\end{multline}
Then denoting $\Sigma \mathbf{.} = ._{LL} + ._{LR} + ._{RL} + ._{RR}$ and finding unknown cut-on frequencies $\omega_A$ for which $\lambda = 1$ leads to the eigenproblem:
\begin{equation}
 \left[ 
\begin{array}{cc} 
 \Sigma \mathbf{K}^s & -\Sigma \mathbf{C} \\
 \mathbf{0} & \Sigma \mathbf{K}^a \end{array} \right] \left\lbrace  \begin{array}{c}
 \mathbf{u}\\
 \mathbf{p}
 \end{array} 
 \right\rbrace = \omega_A^2\left[ 
\begin{array}{cc}
 \Sigma \mathbf{M}^s & \mathbf{0} \\
 \rho_0 \Sigma \mathbf{C}^T & \Sigma \mathbf{M}^a
 \end{array} 
 \right] \left\lbrace 
\begin{array}{c}
 \mathbf{u}\\
 \mathbf{p}
 \end{array} 
 \right\rbrace 
\label{appearance_elasto}
\end{equation}
Henceforth, the generalization of the reduced formulation is straightforward and leads to the eigenproblem:
\begin{equation}
 \left[\begin{array}{cc}
  -\mathbf{\mathbb{K}}_{RL} & -(\mathbf{\mathbb{K}}_{LL}+\mathbf{\mathbb{K}}_{RR}) \\
  \mathbf{0} & -\mathbf{\mathbb{K}}_{RL}
 \end{array}\right]
 \left\lbrace \begin{array}{c}
  \boldsymbol{\mu}_L \\
  \boldsymbol{\mu}_R
 \end{array}\right\rbrace 
 = \lambda \left[\begin{array}{cc}
  \mathbf{0} & \mathbf{\mathbb{K}}_{LR} \\
  -\mathbf{\mathbb{K}}_{RL} & \mathbf{0}
  \end{array}\right] 
 \left\lbrace \begin{array}{c}
  \boldsymbol{\mu}_L \\
  \boldsymbol{\mu}_R
 \end{array}\right\rbrace 
\end{equation}
with the dynamic stiffness operator \cite{Ohayon1992}:
\begin{equation}
\mathbb{K} = -\omega^2 \left[ \begin{array}{cc}
\mathbf{M}^s & \mathbf{0} \\
\mathbf{0} & \mathbf{M}^a
\end{array} \right] 
 + j \omega
 \left[ \begin{array}{cc}
\mathbf{0} & \rho_0 \mathbf{C} \\
 \rho_0 \mathbf{C}^T & \mathbf{0}
\end{array} \right] 
+ \left[ \begin{array}{cc}
\mathbf{K}^s & \mathbf{0} \\
\mathbf{0} & \mathbf{K}^a
\end{array} \right] 
\end{equation}
Solution $\mu$ defined Eq. \ref{reducedeig} is written:
\begin{equation}
\mu = \left\lbrace 
\begin{array}{c}
 \mathbf{u}\\
 \mathbf{\Psi}
 \end{array} 
 \right\rbrace 
\end{equation}
where $\mathbf{u}$ stands for the spatial displacement and $\mathbf{\Psi}$ is the acoustic velocity potential.

\end{document}